# A Fuzzy Control Algorithm for the Electronic Stability Program optimized for tyre burst control

Luca Piancastelli, Leonardo Frizziero, Simone Marcoppido

*DIEM, University of Bologna*
*viale Risorgimento, 2 - 40136 Bologna, Italy*
*e-mail:* [*luca.piancastelli@unibo.it*](mailto:luca.piancastelli@unibo.it)

Eugenio Pezzuti

*University of Rome Tor Vergata*
*Faculty of Engineering*
*via del Politecnico, 1 – 00133 Rome, Italy*
*e-mail:* [*pezzuti@mec.uniroma2.it*](mailto:pezzuti@mec.uniroma2.it)

## ABSTRACT

This paper introduces an improved Electronic Stability Program for cars that can deal with the sudden burst of a tyre. The Improved Electronic Stability Program (IESP) is based on a fuzzy logic algorithm. The IESP collects data from the same sensors of a standard ESP and acts on brakes/throttle with the same actuators. The IESP reads the driver steering angle and the dynamic condition of the car and selectively acts on throttle and brakes in order to put the car on the required direction even during a tyre burst.

**Key words:** ESP; Fuzzy Logic; Tyre road contact; ABS.
**Thematic group:** Innovative methods in industrial design.

## 1. INTRODUCTION

The IESP is an active safety device conceived to reduce the probability of an accident. The IESP improves car stability during normal driving and achieves maximum safety during a tyre burst. Tyre deflation and burs are caused by:
• an impact with sufficient energy to cause serious damage to tyre structure ;
• age;
• puncture.

The simulations described herein consider a very rapid tyre deflation that cannot be dealt by a standard driver.

## 2. THE MATH MODEL OF THE CAR [8]

A lumped mass mathematical model of the car was implemented in order to simulate the behaviour of the ESP system. In this model the car frame and the tires were considered infinitely stiff. Wheel assembly masses where ignored and it was assumed that the resultant of aerodynamic loads passes by the car Gravity Center (GC). Car body has 6 Degrees of Freedom (DOF) defined by GC coordinates and Euler's angles (pitch, roll and yaw). Each wheel has two additional DOFs: rotation and vertical suspension movement. The resulting non linear equations are solved with numerical integration [5], [8].

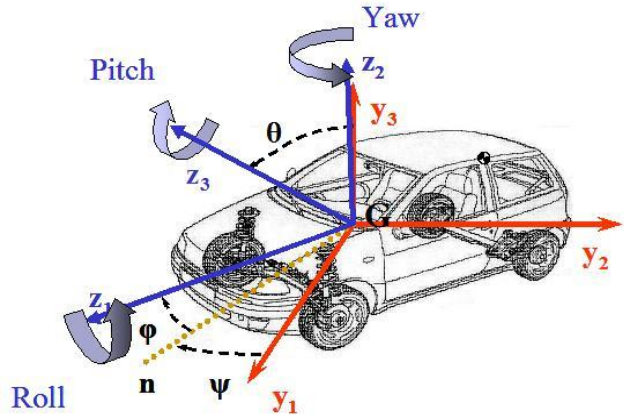

**Figure 1. reference systems**

### 2.1. TYRE-ROAD CONTACT MODEL

Tyre road contact model should be accurate in order to obtain realistic results from simulation. Two distinct models were introduced in the IESP software for correctly inflated and under inflated tyre [5].

TYRE-ROAD MODEL FOR CORRECTLY INFLATED TYRES – It has been modelled through a semi empirical algoritm starting from tyre longitudinal $\mu_{long}$ and transversal friction $\mu_{trasv}$. By changing these two data it is possible to change road or tyre type and condition. $\mu_{long}$ and $\mu_{trasv}$ are calculated starting from the diagrams depicted in Figures 2 and 3. For slip values $\sigma$ inferior to $\sigma_p$, the two friction coefficient are calculated by the formula:

$$\left(\frac{\mu_{trasv}}{(\mu_{trasv})_{max}}\right)^2 + \left(\frac{\mu_{long}}{(\mu_{long})_{max}}\right)^2 = 1 \tag{1}$$

For slippage near or superior the maximum the formula (1) is excessively approximate. In this case an algorithm was developed to correct this problem for values of $\sigma/\sigma_p$ close to unity. This corrective algorithm takes into account the saturation of $\mu_{trasv}$ for $\sigma \geq \sigma_p$. Figures 4 and 5 depict $\mu_{long}$ and $\mu_{trasv}$ as function of $\sigma$ and $\alpha$ (slip angle).

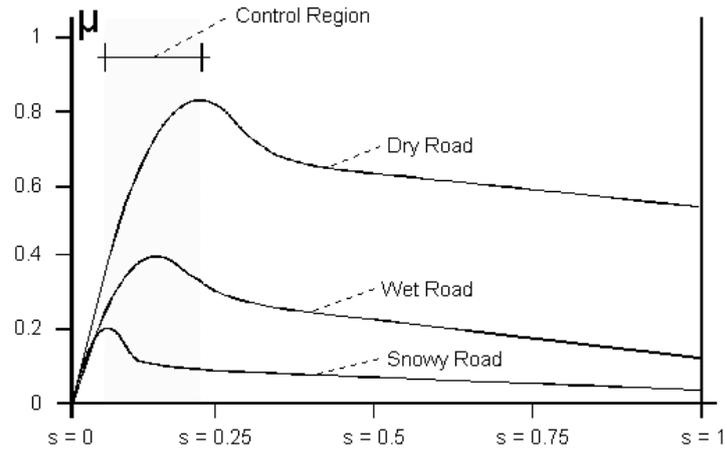

**Figure 2. Flong-σ for different road conditions.**

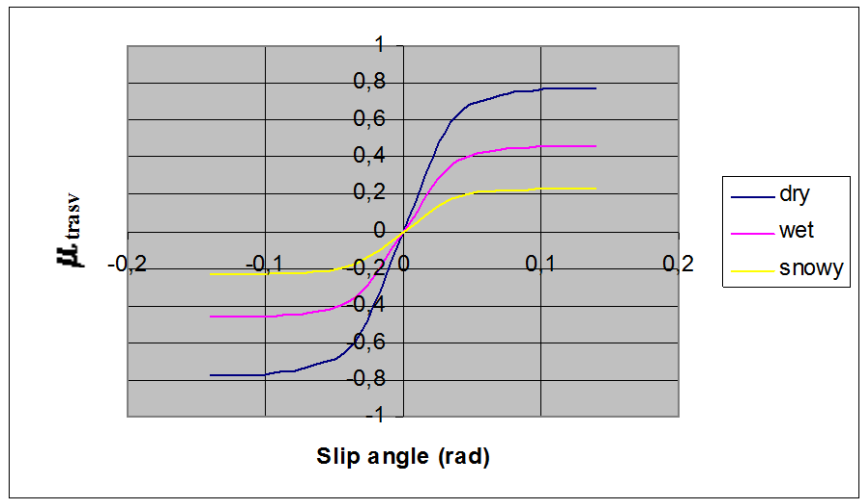

**Figure 3. Ftrasv-α for different road conditions**

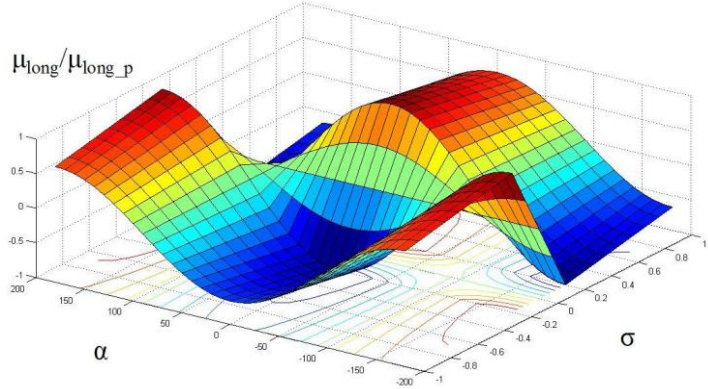

**Figure 4. $\mu_{long}$ normalized for $\sigma$ and $\alpha$ as used in the simulation.**

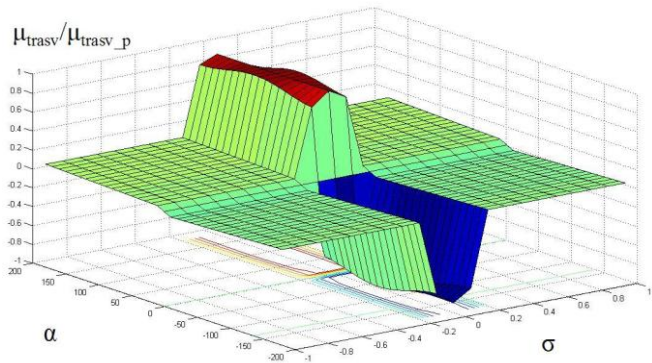

**Figure 5. $\mu_{trasv}$ normalized for $\sigma$ and $\alpha$ as used in the simulation.**

### 2.1.1. TYRE-ROAD CONTACT MODEL FOR UNDER INFLATED TYRES [11], [13]

In this case the friction force T is defined as follows:

$$\vec{T} = \mu(\alpha) \cdot \left| \vec{N} \right|$$

(2)

N is the vertical load on the tyre and T has a direction opposite to the relative speed of tyre and road. In this case the friction coefficient depends only on slip angle. Friction has an elliptic shape as shown is figure 6. In the simulation the deflation process can be introduced in several ways. In the simulation only very rapid complete deflation in 3 s were considered.

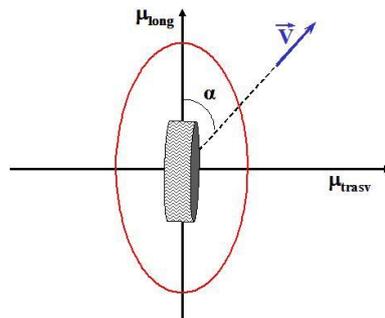

**Figure 6. Polar diagram of friction coefficient for deflated tyre road contact.**

## 3. THE FUZZY AUTOMATIC PILOT

In general, Fuzzy logic differs from conventional logical systems in that it aims at providing a model for approximate rather than precise reasoning. The fuzzy logic, FL, has the Following principal features. (a) The truth-values of FL are fuzzy subsets of the unit interval carrying labels such as true, very true, not very true, false, more or less true, etc.; (b) The truth-values of FL are structured in the sense that they may be generated by a grammar and interpreted by a semantic rule; (c) FL is a local logic in that, in FL, the truth-values as well as the connectives such as and, or, if... then have a variable rather than fixed meaning; and (d) The rules of inference in FL are approximate rather than exact.

The central concept in FL is that of a fuzzy restriction, by which is meant a fuzzy relation which acts as an elastic constraint on the values that may be assigned to a variable. Thus, a fuzzy proposition such as 'Nina is young' translates into a relational assignment equation of the form R(Age (Nina)) = young in which Age (Nina) is a variable, R(Age(Nina)) is a fuzzy restriction on the values of Age(Nina), and young is a fuzzy unary relation which is assigned as a value to R (Age (Nina)) [1], [2].

Moreover, one of the most attractive features of fuzzy set theory is to provide a mathematical setting for the integration of subjective categories represented by membership functions. Indeed, a body of aggregation operations is already available, which may be useful in decision analysis, quantitative psychology and information processing [4].

For the simulations a fuzzy logic automatic pilot was introduce [6], [9], [10], [17], [18]. It controls the steering wheel and the clutch pedal. The brake pedal, throttle pedal position should be given in input since they are not controlled by the autopilot. Required trajectory is defined as a sequence of four parametric bends. The auto pilot is defined by two controllers that act in parallel: the fuzzy logic pilot and the kinematic pilot. The kinematic pilot defines a steering wheel position starting only from the required trajectory The kinematic steering angle δ is defined by the following formula:

$$\delta = \arctan\left(\frac{l}{\rho}\right) \qquad (3)$$

where V is the GC horizontal velocity, l is the wheelbase and ρ is the curvature radius of the trajectory. The fuzzy controller uses a feedback based logic that takes into account of GC position, yaw angle $\psi$, yaw rate $\psi\,'$ and GC velocity V.

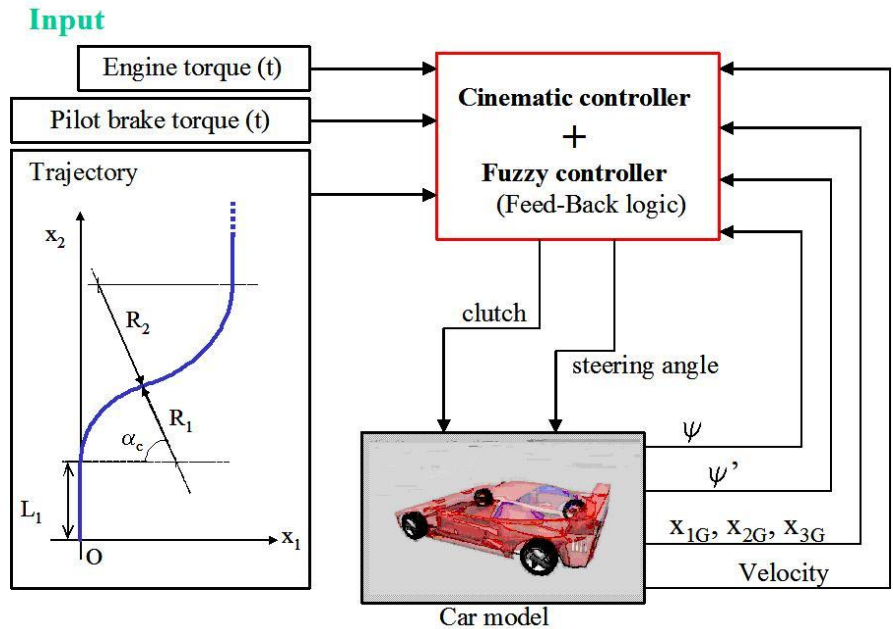

**Figure 7. Autopilot conceptual scheme**

## 4. THE ABS-IESP [7], [11]

The priorities of the integrated ABS-IESP system are the following:

      • Avoid wheel blockage
      • Keep a correct slip angle $\beta$
      • Keep a correct yaw angle $\psi$ '

      To avoid wheel blockage is the first condition since this fact induces absence of directional control or stability if front or rear wheels are concerned  The Anti Blockage System is the ABS. IESP outputs are sent to the ABS in order to avoid wheels blockage. The second priority is to control the slip angle in order to assure to the driver and adequate manoeuvrability and steering efficiency. For this purpose the IESP should monitor the slip angle $\beta$.

The slip angle $\beta$ is defined as the angle between CG velocity vector and vehicle longitudinal axis (see figure 8).
When $\beta$ is large the driver feels in danger since steering tends to lose efficiency. For dry ashpalt in good condition the phisical limit of $\beta$ is about 12°, while for icy road this limit value is reduced to 2°. When this angle is reached the vehicles enters into a spin whatever is the steering wheel position and driving control is lost. An average car driver has no driving experience for $\beta$ angles above  2°.

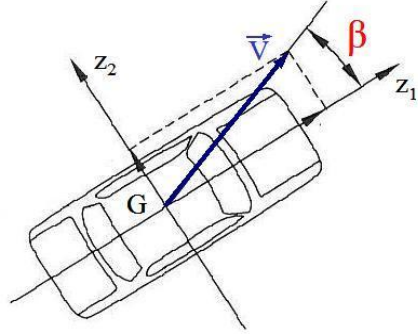

**Figure 8. Slip angle of the car $\beta$**

At last the IESP should control the yaw speed in order to fulfil driver input and avoid adherence limit.

***THE ABS*** [7], [12], [14], [15], [16] – The ABS implemented in this program is subdivided into two modules: the brake modulator and the brake distributor (figure 9). The brake modulator is based on fuzzy logic. Input variables are brake pedal position, IESP output, the slip values $\sigma$ of wheels and their slip rate $\sigma$'. For sake of simplicity the real slip was used instead of the estimated value as in real ABS. The brake distributor takes the four values (one for each wheel) of the brake modulator and the longitudinal and traverse acceleration of the GC. These acceleration are necessary to calculate the vertical load on each wheel:

$$\begin{cases} F_{v\_Fl} = F_{v\_F(s)} + \Delta F_v + \Delta F_{v\_F\_roll} \\ F_{v\_Fr} = F_{v\_F(s)} + \Delta F_v - \Delta F_{v\_F\_roll} \\ F_{v\_Rl} = F_{v\_R(s)} - \Delta F_v + \Delta F_{v\_R\_roll} \\ F_{v\_Rr} = F_{v\_R(s)} - \Delta F_v - \Delta F_{v\_R\_roll} \end{cases} \qquad (4)$$

$$\Delta F_v = \frac{m \cdot a_{long} \cdot h_G}{l} \qquad (5)$$

$$\Delta F_{v\_roll} = \left( \frac{m}{c} \cdot h_G \cdot a_{trasv} \right) \qquad (6)$$

$$\Delta F_{v\_R\_roll} = \frac{\Delta F_{v\_roll}}{1 + \dfrac{K_{rollF}}{K_{rollR}}} \qquad (7)$$

$$\Delta F_{v\_F\_roll} = \Delta F_{v\_roll} - \Delta F_{v\_R\_roll} \qquad (8)$$

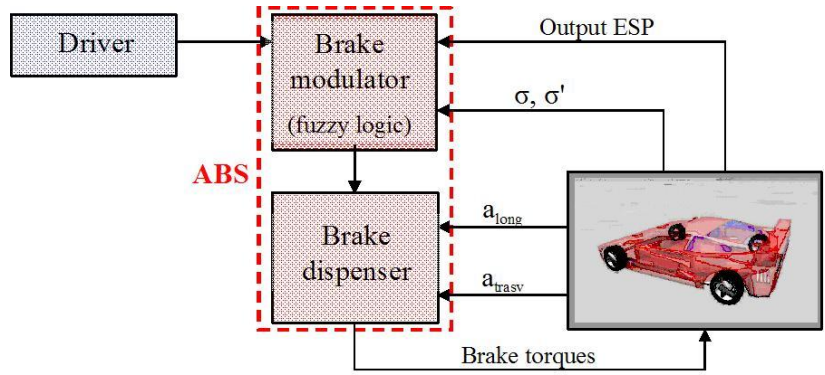

**Figure 9. ABS conceptual scheme**

***THE IESP*** [7], [12] – The IESP is subdivided in four distinct block: the "the monotrack steady state", the "the slip angle estimator", the "fuzzy controller" and the "yaw moment dispenser" (figure 10). The logical intervention steps of the IESP are the following:

1. IESP reads the sensors and defines vehicle condition and slip (actual dynamic response);
2. at the same time IESP evaluates the ideal vehicle response;
3. then IESP compares the ideal and the actual response and outputs the optimal command to minimize the difference of the two responses.

The "actual" can be calculated, for the stability control, from the slip angle $\beta$ e the yaw rate $\psi'$. $\psi'$ can be directly measured from an ad hoc sensor. More complex is the determination of the slip angle $\beta$, since a commercial sensor for this value in not available.

The slip angle $\beta$ can be approximately derived from other measured data and the application of kinematic and dynamic formulas.

For this purpose a virtual model of the vehicle has been implemented.

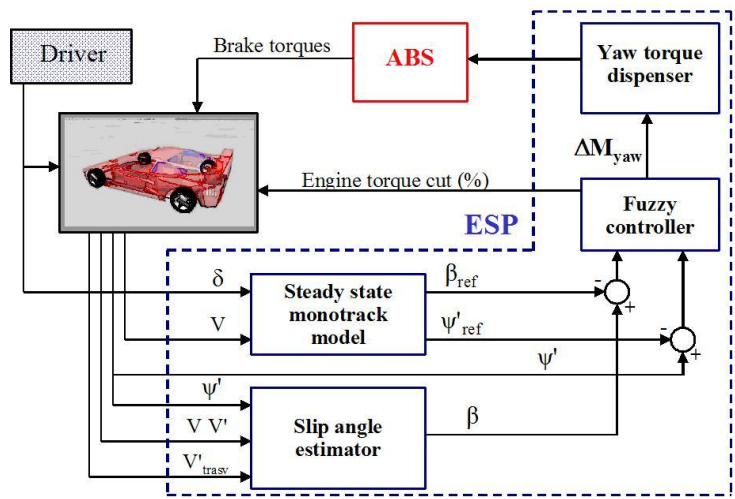

**Figure 10. IESP conceptual scheme**

The "ideal response" is the vehicle behaviour that the car should have in theoretical ideal condition (infinite friction and no mass). In this condition the driver does not have to implement any correction. Refence values for the ideal yaw rate $\psi_{ref}$' and the ideal slip angle $\beta_{ref}$ are calculated by the monotrack model that used a linear steady state model. As it can be seen in figure 11, the ideal yaw rate $\psi_{ref}$' can be obtained by the formula:

$$\dot{\psi}_{ref} = \frac{V_x}{\rho} \cong \frac{\tan\delta}{l} \cdot V \qquad (9)$$

where V is the GC horizontal velocity, δ is steering wheel angle, l is the wheelbase and ρ is the curvature radius of the trajectory.

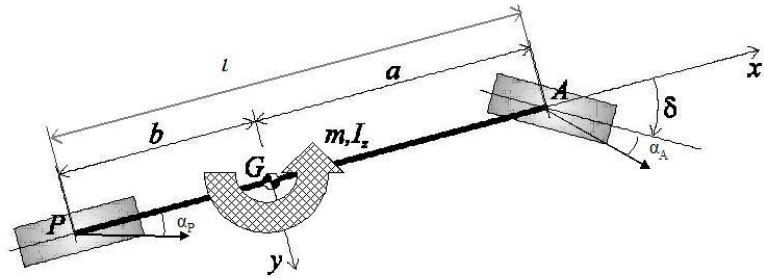

**Figure 11. Single-track model of the vehicle**

The vehicle has a neutral behaviour is $\psi_{ref}$' is close to the effective yaw rate $\psi$'. This makes the vehicle unstable for increasing speeds. In order to keep stability the (9) has been modified:

$$\dot{\psi}_{ref} = \frac{\tan\delta}{l \cdot \left(1 + k_{us} \cdot V^2\right)} \cdot V \qquad (10)$$

where $k_{us}$ è is a positive constant: an increase in $k_{us}$ values makes the car more understeering and more stable. Similar considerations can be made on the slip angle:

$$\beta_{ref} = \frac{b}{\rho} - \alpha_p \cong \frac{\tan\delta}{l \cdot \left(1 + k_{us} \cdot V^2\right)} \cdot \left(b - k_{ps} \cdot V^2\right) \qquad (11)$$

where $\alpha_p$ is the derive angle of the rear wheel and $k_{ps}$ is a coefficient that, like in (10), increases the understeering behaviour of the car. Formulas (10) and (11) are purely kinematic and do not take into account friction. In fact when a vehicle runs a corner with low attrition or excessive speed it is possible that the centrifugal force overcomes the resultant of the tyre-soil friction force with a loss of the desired trajectory. A vehicle that travels on a corner with curvature radium ρ in steady state condition has the following lateral acceleration:

$$\dot{V}_y = \frac{V_x^2}{\rho} = \dot{\psi} \cdot V_x \tag{12}$$

In order to keep the path, the lateral acceleration should not overcome the transversal tyre-road friction value:

$$\dot{V}_y \leq \mu_{y\_max} \cdot g \tag{13}$$

this fact implies a maximum value for the yaw rate:

$$\dot{\psi}_{limit} \leq \frac{\mu_{y\_max} \cdot g}{V_x} \cong \frac{\mu_{y\_max} \cdot g}{V} \tag{14}$$

The reference value for the IESP is the lowest value among those calculated through formulas from (9) to (14).

The "fuzzy controller" block has in input $\beta$, $\beta_{ref}$, $\psi$' and $\psi_{ref}$' and it is composed by two fuzzy controllers: "$\Delta M_{yaw}$ fuzzy logic" and "Engine torque cut fuzzy logic". The input variables of the first controller are: slip angle error (BetaErr, $e_\beta$), calculated from the following formula:

$$e_\beta = \beta - \beta_{ref} \tag{15}$$

and the yaw angle error (PsidotErr),

$$e_{\dot{\psi}} = \dot{\psi} - \min\left(\dot{\psi}_{ref}; \dot{\psi}_{limit}\right) \tag{16}$$

The output variable is the corrective yaw moment (DeltaMyaw). The membership functions for BetaErr are: negative large (-B), negative small (-S), null (Z), positivo small (S), large (B).

Analogous membership functions (-B), (S), (Z), (S), (B) are chosen for (PsidotErr).

Nine membership function are chosen for the output variable DeltaMyaw.

In some cases it is impossibile to minimize at the same time the yaw rate error (PsidotErr) and the slip rate error (BetaErr). In this cases the controller miminizes BetaErr first. IESP privileges stability to path, in fact once lost the stability it would have been anyway impossibile to keep the path. In the case as the stability is restored ($\beta$), IESP correct path ($\psi$').

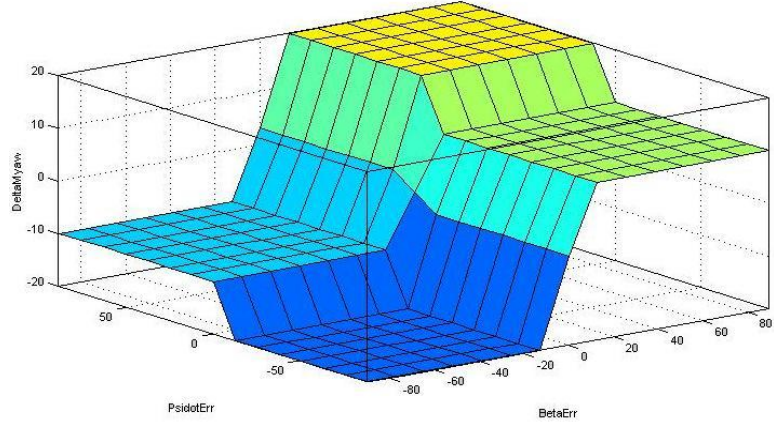

**Figure 12. Graphic visualization of the correlation of output variable "DeltaMyaw" and the input variables "BetaErr" e "PsidotErr"**

The input variables of the second controller "Engine torque cut fuzzy logic" are: the amount of overcome of the adherence limit (Limit), given by:

$$Limit = \dot{\psi}_{ref} - \dot{\psi}_{limit} \tag{17}$$

where $\psi_{ref}$' e $\psi_{limit}$' are calculated by (10) and (14). The output variable is the percent of torque cut (enginetorquecut). Membership functions for Limit are: null (N), small (S), medium (M), Large (B) (see table 1).

| Name of membership function | Numeric Value (constant) (%) | Comment |
|---|---|---|
| Z | 0 | Zero |
| S | 30 | Small |
| M | 60 | Medium |
| B | 95 | Large |

**Table 1. Membership functions of enginetorquecut.**

In this way the IESP cut progressively the engine torque as the adherence limit is overcome. The "Yaw torque dispenser" starts from the corrective yaw moment required (DeltaMyaw, $\Delta M_{yaw}$) and outputs the amount of corrective brake to be applied on each wheel. These output values are added to the brake position and inputed to the ABS. The subdivision of the Yaw moment between the front $M_F$ and rear $M_R$ depends on the vertical load on the two axis:

$$M_F = \frac{F_{v\_Fl} + F_{v\_Fr}}{F_{v\_tot}} \cdot \Delta M_{yaw} \tag{18}$$

$$M_R = \frac{F_{v\_Rl} + F_{v\_Rr}}{F_{v\_tot}} \cdot \Delta M_{yaw} \tag{19}$$

Where:

$$F_{v\_tot} = F_{v\_Asx} + F_{v\_Adx} + F_{v\_Psx} + F_{v\_Pdx} \tag{20}$$

is the sum of the vertical loads on wheels  The variations of brake moments to be applied are then:

$$\Delta M_{brake\_F} = \frac{2 \cdot r_w}{c_F} \cdot M_F \tag{21}$$

$$\Delta M_{brake\_R} = \frac{2 \cdot r_w}{c_R} \cdot M_R \tag{22}$$

where $r_w$ is wheel radius $c_F$ are $c_R$ are half the width of the front and  rear  axis respectively.

On the front axes the brake moment is applied on only one wheel. If $\Delta M_{yaw\_F}$   is positive/negative then  brake is applied on right/left wheel. On the rear axis the moment $\Delta M_{yaw\_R}$ is applied on both wheels with opposite sign in order to obtain the required value.

## 5. RESULTS ESP

Two different simulation where performed to test the system:

- Tests with tyre burst on straight roads and successive manoeuvre of stop aside.
- Tests with tyre burst on corners

In both cases several parameters where varied:

- Vehicle velocità and engine maximum torque;
- Burst wheel (front left, front right, rear left, rear right;
- longitudinal ($\mu_{long}$) burst and transversal ($\mu_{trasv}$)$_{burst}$ friction coefficient of the damaged tyre;
- with or without IESP;
- trajectory curvature: null, 50 m, 100 m.

or the undamaged tyres have the same for all the simulations ($\mu_{long\_p} = \mu_{trasv\_p} = 0.9$). ($\mu_{long}$)$_{burst}$ and ($\mu_{trasv}$)$_{burst}$, are widely varied from 0.8 to 0.05 since in this case no data have been found. In this way it was possible to evaluate:

- trajectory error;
- slip angle and slip angle error;
- yaw rate error;

AN EXAMPLE OF ESP APPLICATION − A right bend with curvature radius of 100 m, initial velocity 95 Km/h, $(\mu_{long})_{burst} = (\mu_{trasv})_{burst} = 0.05$. Right rear tyre bursts 6 s from the beginning of the bend with $(\mu_{long})_{burst} = (\mu_{trasv})_{burst} = 0.05$. Spin occurs without IESP. Figure 13 depicts the slip angle vs. time. It is possibile to see that from t = 6s (burst time) $\beta$ grows from 0° to about 700° (two spins). The same test wit the IESP (figure 14) shows tha $\beta$ is always under 2.5°. The vehicle remains stable and the the maximum trajectory error is 0.8 m. (see figure 15).

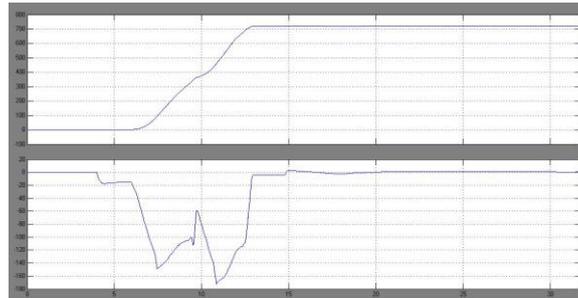

**Figure 13.** $\beta$ (up, degree) $\psi'$ (bottom, degree/s) versus time (s), 100 m radius bend, 95 km/h, IESP off, tyre burst at 6 s

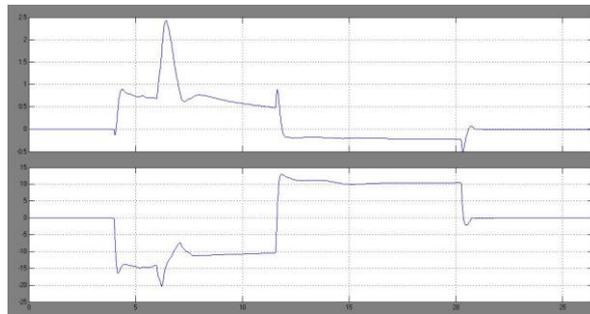

**Figura 14.** $\beta$ (up, degree) $\psi'$ (bottom, degree/s) versus time (s), 100 m radius bend, 95 km/h, IESP on, tyre burst at 6 s

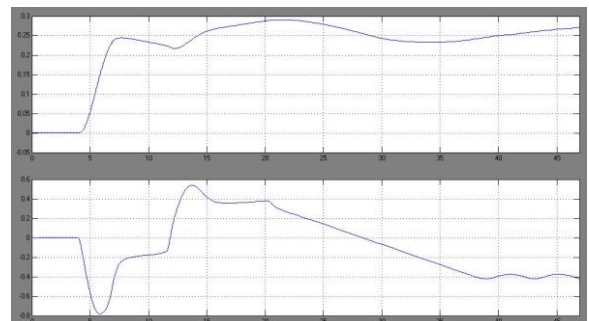

**Figure 15** medium(up, m) and instantaneous (bottom, m) versus time (s), 100 m radius bend, 95 km/h, IESP on, tyre burst at 6 s.

## 6. CONCLUSIONS

This paper demonstrates that it is possible to design an electronic stability system (IESP) for trajectory and stability control that can work even during a tyre burst:

- Damage of front tyres: IESP can avoid spin and halves the trajectory error. In this case impact with road side protections cannot be avoided but injures can be significantly reduced or avoided.
- Damage of rear tyres: at low speed a rear tyre burst is a situation that can be controlled even by and average driver. At high speed spin is unavoidable even for expert drivers. In this case IESP avoids the spin and contains the trajectory error within 1 m for speed sbelowl 250 km/h.

IESP can solve safety problem for rear wheel burst and can significantly improve safety for front tyre bursts.

The prescriptive approach illustrates what needs to be done to advance beyond a simply descriptive system. What is desired is that such an approach should appear naturally within a suitably improved fuzzy logic theory itself [3].

## 8. USED SYMBOLS

$\psi$: yaw angle [rad]

θ: pitch angle [rad]

φ: roll angle[rad]

m: vehicle mass [kg]

β: slip angle [rad]

$F_{long}$: longitudinal force on tyre [N]

$F_{trasv}$: transversal force on tyre [N]

N, $F_v$: vertical force on tyre [N]

σ: longitudinal slip [-]

$σ_M$: longitudinal slip with maximum longitudinal friction coefficient [-]

α: derive angle [rad]

l: wheelbase [m]

b: distance GC to rear axes [m]

$c_F$: front axes length [m]

$c_R$: rearaxes length [m]

$r_w$: wheel radius [m]

$μ_{long}$: longitudinal friction coefficient [-]

$μ_{trasv}$: transversal friction coefficient [-]

t: time [s]

δ: steering angle [rad]

ρ: trajectory curvature radius [m]

$K_{us}$, $K_{ps}$: IESP internal parameters [rad sec$^2$ m$^{-2}$]

$V_x$, $V_y$: horizontal velocità of GC along x, y axis [m/s]

$ψ_{ref}$': reference yaw rate [rad/s]

$β_{ref}$: reference slip angle [rad]

$ψ_{limit}$': adherence limit yaw rate [rad/s]

$e_{ψ'}$,: yaw rate error [rad/s]

$e_β$: slip angle error [rad]

$\Delta M_{yaw}$: corrective yaw moment [N m]

$M_F$, $M_R$: corrective yaw moment on front (F) and rear (R) axis [N m]

$M_{Brake}$: Brake moment on wheel [N m]